\documentclass[twocolumn]{article} 
\usepackage{graphicx} 

\long\def\comment#1{}

\makeatletter
\def\@normalsize{\@setsize\normalsize{10pt}\xpt\@xpt
\abovedisplayskip 10pt plus2pt minus5pt\belowdisplayskip
\abovedisplayskip \abovedisplayshortskip \z@
plus3pt\belowdisplayshortskip 6pt plus3pt
minus3pt\let\@listi\@listI}

\def\subsize{\@setsize\subsize{12pt}\xipt\@xipt}
\def\section{\@startsection {section}{1}{\z@}{1.0ex plus
1ex minus .2ex}{.2ex plus .2ex}{\large\bf}}
\def\subsection{\@startsection
   {subsection}{2}{\z@}{.2ex plus 1ex} {.2ex plus .2ex}{\subsize\bf}}
\makeatother

\begin{document}

\date{}

\title{\huge \bf {On Subgraph Isomorphism}}

\author{Sergey Gubin
 \thanks{sgubin@genesyslab.com
 }
}

\maketitle
\thispagestyle{empty}


{\hspace{1pc} {\it{\small Abstract}}{\bf{\small---Article explicitly expresses Subgraph Isomorphism by a polynomial size asymmetric linear system.

 }}
 }


\section*{Introduction}
\label{s:intro}

$\mbox{	}$ In 1988, Yannakakis proved \cite{yan} that the Traveling Salesman Problem's (TSP) polytope cannot be expressed by a polynomial size symmetric linear program, where symmetry means that the polytope is an invariant under node relabeling. Because TSP is a NP-complete problem \cite{karp}, the theorem holds for all NP-complete problems. The question about the size of asymmetric linear models was left open in \cite{yan} and it has remained open since.
\newline\indent
$\mbox{	}$ This article answers that question. We present an explicit polynomial size asymmetric linear model for Subgraph Isomorphism (SubGI). Since SubGI is a NP-complete problem \cite{cook0}, this result is complimentary to the Yannakakis theorem.
\newline\indent
$\mbox{	}$ The polynomial size asymmetric linear system is built based on an arbitrary but fixed labeling of graphs involved -  hence the system's asymmetry. The polynomial size for the system is achieved by immersing the problem in a space of higher dimension, where variables present relabeling possibilities for vertex couples.
\newline\indent
$\mbox{	}$ We illustrate our method with several examples. Particularly, we explicitly present polynomial size asymmetric linear programs for TSP and for the Satisfiability Problem for conjunctive normal forms (SAT). 

\section{Subgraph Isomorphism}
\label{s:subgi}

$\mbox{	}$ Let $G$ be a given graph - we will call it an input. Let $S$ be another given graph - we will call it a pattern. The problem is whether $G$ contains a subgraph which is isomorphic to $S$. For any given couple of graphs $(G,S)$, this decision problem is a SubGI instance. Its size can be estimated by the number of vertices in graph $G$.
\newline\indent
$\mbox{	}$ Any graph may be seen as a relation. So, SubGI may be seen as a finite version of the following general problem: whether a given relation posseses a given property. That explains the theoretical and practical importance of SubGI.
\newline\indent
$\mbox{	}$ SubGI is a NP-complete problem \cite{cook0}. The Ullmann algorithm \cite{ullmann} is the best known method to solve the problem. Yet it and other known general methods are inefficient. Up to date, the efficient methods were known only for particular types of graph couples $(G,S)$ \cite[and others]{hopcroft, luks, david}. 
\newline\indent
$\mbox{	}$ This article describes a reduction of SubGI to a system of linear equations and inequalities. The reduction's computational complexity and the resulting system's size are polynomial over the size of SubGI. For a given couple of graphs $(G,S)$, the resulting system has solutions iff input $G$ contains a subgraph isomorphic to pattern $S$.
\newline\indent
$\mbox{	}$ As well as for graphs, our reduction works for (multi) digraphs with (multi) loops. We will present the reduction for the multi digraph version of SubGI which is, in many cases, more practical. So, input $G$ and pattern $S$ are (multi) digraphs with (multi) loops everywhere below in this article.
\newline\indent
$\mbox{	}$ Because our system contains a polynomial number of linear equations and inequalities with a polynomial number of unknowns, it can be solved in polynomial time by, for example, the Khachiyan ellipsoid algorithm \cite{khach, gls}.

\section{Base polytope}
\label{s:base}

$\mbox{	}$ Let $n$ be a natural number. Let the following variables be unknowns:
\[
\begin{array}{rlll}
x_{ij \mu\nu} = x_{ji \nu\mu}: & i,j,\mu,\nu = 1,2,\ldots,n & i \neq j & \mu \neq \nu \\
y_{ii \mu\mu} = y_{jj \nu\nu}: & i,j,\mu,\nu = 1,2,\ldots,n \\
\end{array}
\]
In the case of $n=1$, variables $x_{ij \mu\nu}$ are missing indeed.
\newline\indent
$\mbox{	}$ Let's consider the following linear system:
\begin{equation}
\label{e:base}
\left \{ \begin{array} {l}
x_{ij \mu\nu} = x_{ji \nu\mu}, ~ x_{ij \mu\nu} \geq 0 \\
\mbox{- where } i,j,\mu,\nu = 1,2,\ldots,n, ~ i \neq j, ~ \mu \neq \nu \\
\\
\sum_{\mu = 1,~\mu\neq\nu}^n x_{ij \mu\nu} = y_{jj \nu\nu}, \\
\mbox{- where } i,j,\nu = 1,2,\ldots,n,  ~ i \neq j \\
\\
\sum_{i = 1,~i \neq j}^n x_{ij \mu\nu} = y_{jj \nu\nu}, \\
\mbox{- where } j,\mu,\nu = 1,2,\ldots,n,  ~ \mu \neq \nu \\
\\
\sum_{\nu = 1}^n y_{jj \nu\nu} = 1, ~ y_{jj \nu\nu} \geq 0 \\
\mbox{- where } j = 1,2,\ldots,n \\
\end{array} \right.
\end{equation}
The system can be described with the following box matrix of size $n\times n$:
\[
B = \left ( \begin{array}{cccc}
Y_{1,1} & X_{1,2} & \ldots & X_{1,n} \\
X_{2,1} & Y_{2,2} & \ddots & X_{2,n} \\
\vdots & \ddots & \ddots & \vdots \\
X_{n,1} & X_{n,2} & \ldots & Y_{n,n} \\
\end{array} \right )_{n\times n}
\]
The $i$-th diagonal box in box matrix $B$ is the following diagonal matrix:
\[
Y_{ii} = \mbox{diag}(y_{i,i,1,1}, ~ y_{i,i,2,2},~\ldots,~ y_{i,i, \nu,\nu},~\ldots,~ y_{i,i,n,n})
\]
The $(i,j)$-th off-diagonal box in box matrix $B$ is the following matrix:
\[
X_{ij} = \left( \begin{array}{cccc}
0&x_{i,j,1,2}&\ldots&x_{i,j,1,n} \\
x_{i,j,2,1}&0&\ddots&x_{i,j,2,n} \\
\vdots&\ddots&\ddots&\vdots \\
x_{i,j,n,1}&x_{i,j,n,2}&\ldots&0 \\
\end{array}\right )_{n\times n}
\]
System \ref{e:base} reflects the following relations between elements of box matrix $B$: 
\newline
(1) $B$ is a symmetric matrix: $X_{ij} = X_{ji}^T$;
\newline
(2) The total of each column in matrix $X_{ij}$ does not depend on $i$ but only on box column $j$ and on the column in this box column. The total is the appropriate element in diagonal matrix $Y_{jj}$;
\newline
(3) The total over $i$ of elements $x_{ij \mu\nu}$ does not depend on $\mu$ but only on box column $j$ and on the column $\nu$ in this box column. This total is the appropriate element in matrix $Y_{jj}$ - element $y_{jj \nu\nu}$;
\newline
(4) The $(j,\nu)$-th columns in off-diagonal boxes $X_{ij}$ of box matrix $B$ constitute a doubly stochastic matrix multiplied by element $y_{jj\nu\nu}$.
\newline 
(5) The total of all elements in matrix $Y_{jj}$ is equal $1$;
\newline
(6) Due to the matrix's symmetry, all of the above is true in the horizontal direction, too;
\newline\indent
$\mbox{	}$ System \ref{e:base} always has solutions. The following solution is minimal in the sense of Euclidean norm - we call it a \emph{center}:
\[
x_{ij \mu\nu} \equiv \frac{1}{n(n-1)}, ~ y_{jj\nu\nu} \equiv \frac{1}{n}
\]
Obviously, the set of all solutions of system \ref{e:base} is a convex set. Also, because system \ref{e:base} is a linear system, the set is a polytope. We call this polytope a \emph{base polytope}. 
\newline\indent
$\mbox{	}$ The following solution of system \ref{e:base} is a vertex of the base polytope: there is one and only one non-zero element in each box $Y_{ii}$ and $X_{ij}$. Obviously, all non-zero elements in the boxes are equal $1$ and they are arranged in a grid of elements in matrix $B$, one element per box. We call any such solution of system \ref{e:base} a \emph{solution grid}.
\newline\indent
$\mbox{	}$ The following lemma shows that all vertices of the base polytope are solution grids.
\begin{description}
\item[Lemma 1]
Any solution of system \ref{e:base} is a convex combination of solution grids.
\item[Proof]
System \ref{e:base} consists of linear equations and the following inequalities:
\[
 x_{ij \mu\nu} \geq 0, ~ y_{jj\nu\nu} \geq 0
\]
- where all indexes are in their appropriate ranges. The linear equations have solutions - the center, for example. The solutions constitute a linear subspace in the linear space of all $n^2 \times n^2$ matrices with real elements. Thus, vertices of base polytope are those points in the linear subspace where the number of variables which equal $0$ is maximal possible. It so happens that these points are the solution grids, QED.
\end{description}

\section{Compatibility matrix}
\label{s:matrix}

$\mbox{	}$ Let digraphs $G$ and $S$ be the given SubGI instance $(G,S)$. Let $V_G$ and $V_S$ be vertex sets of the input and pattern appropriately. Obviously,
\[
|V_G| \geq |V_S|
\]
- the instance would have resolution ``NO'' otherwise.
\newline\indent
$\mbox{	}$ Now, let's add $|V_G| - |V_S|$ isolated vertices to pattern $S$. Let's preserve notion $S$ for the resulting pattern. Let  $n$ be the number of vertices in the input and the pattern after the addition of isolated vertices:
\[
n = |V_G| = |V_S|
\] 
Obviously, the SubGI instance $(G,S)$ emerging after the addition of isolated vertices has the same resolution as the original instance indeed.
\newline\indent
$\mbox{	}$ Let's arbitrarily label/enumerate vertices in input $G$ and pattern $S$. Let $A_G$ and $A_S$ be the adjacency matrices of the input and pattern appropriate to the labeling. Obviously, SubGI instance $(G,S)$ has resolution ``YES'' iff there exists such a relabeling of pattern $S$ that all elements of matrix $A_S$ emerging after that relabeling will be less than or equal to the appropriate elements of matrix $A_G$. In other words, SubGI instance $(G,S)$ has resolution ``YES'' iff the following integral quadratic system has solutions\footnote{For two matrices $A=(a_{ij})$ and $B=(b_{ij})$ of the same size, relation $A\geq B$ means that
\[
\forall i,j ~ (a_{ij} \geq b_{ij}).
\]
}:
\begin{equation}
\label{e:quadratic}
A_G \geq X A_S X^T
\end{equation}
- where $X$ is the unknown permutation matrix of size $n\times n$. Permutation matrix $X$ presents the unknown vertex relabeling of pattern $S$ after which the existence of an input's subgraph isomorphic to $S$ has to become self-evident. Obviously, such a relabeling of $S$ exists iff $G$ has at least one subgraph isomorphic to $S$.
\newline\indent
$\mbox{	}$ To solve system \ref{e:quadratic}, let's build the following matrix which we call a \emph{compatibility matrix}.
\newline\indent
$\mbox{	}$ Let the input and pattern's adjacency matrices be as follows:
\[
A_G = (g_{\mu\nu})_{n\times n}, ~ A_S = (s_{ij})_{n\times n}
\]
For each couple of pattern's vertices, let's build a \emph{compatibility box}. The compatibility box for vertices with labels $i$ and $j$ is the following matrix $C_{ij} = (e_{ij\mu\nu})_{n\times n}$:
\begin{equation}
\label{e:def}
e_{ij\mu\nu} = \left \{ \begin{array}{cl}
1, & s_{ij} \leq g_{\mu\nu} ~ \wedge ~ s_{ji} \leq g_{\nu\mu}\\
0, & s_{ij} > g_{\mu\nu} ~ \vee ~ s_{ji} > g_{\nu\mu}\\
\end{array} \right.
\end{equation}
Compatibility box $C_{ij}$ shows all possible re-enumerations for the pattern's vertices $i$ and $j$ with disregard to the rest of the pattern's vertices. Obviously, compatibility boxes $C_{ii}$ are diagonal matrices. And all diagonal elements in compatibility boxes $C_{ij}$, $i\neq j$, are equal $0$.
\newline\indent
$\mbox{	}$ The compatibility matrix for SubGI instance $(G,S)$ is the following box matrix:
\[
C = (C_{ij})_{n\times n}
\]
The compatibility matrix aggregates all compatibility boxes in accordance with their indexes.
\newline\indent
$\mbox{	}$ Obviously, integral quadratic system \ref{e:quadratic} has a solution iff in the compatibility matrix there is a grid of elements, one element per compatibility box, in which all elements are equal $1$:
\[
\gamma = \{e_{ij\mu\nu} = 1 ~|~ \mu = \mu(i), ~ \nu = \nu(j)\}
\]
- where $\gamma$ is the grid. Any such grid of elements in compatibility matrix $C$ we call a \emph{solution grid}\footnote{In the next section, we will show that the solution grids from this section and the solution grids from the previous section are the same.}, too.
\begin{description}
\item[Lemma 2]
SubGI instance $(G,S)$ has resolution ``YES'' iff compatibility matrix $C$ contains a solution grid.
\item[Proof]
Any solution grid defines a vertex relabeling of $S$ which satisfies system \ref{e:quadratic}, QED.
\end{description}

\section{Linear model for SubGI}
\label{s:model}

$\mbox{ }$ The similarities between compatibility matrix $C$ and the base polytope $B$ are obvious. Due to lemmas 1 and 2, we can decide about the existence/absence of solution grids in matrix $C$ searching matrix $B$ for solution grids subject to the following constrains:
\begin{equation}
\label{e:cons}
x_{i_0 j_0 \mu_0 \nu_0 } = 0, ~ y_{j_0 j_0 \nu_0 \nu_0} = 0
\end{equation}
- where indexes are the indexes of all those elements of compatibility matrix $C$ which are equal $0$,
\[
e_{i_0 j_0 \mu_0 \nu_0 } = e_{j_0 j_0 \nu_0 \nu_0} = 0
\]
Then, lemmas 1 and 2 imply the following polynomial size asymmetric linear model for SubGI.
\begin{description}
\item[Theorem]
SubGI instance $(G,S)$ has resolution ``YES'' iff the aggregated system \ref{e:base} and \ref{e:cons} has a solution.
\item[Proof]
Any solution of the aggregated system is a convex hull of solution grids. There is a solution grid iff the resolution for instance $(G,S)$ is ``YES'', QED.
\end{description}
$\mbox{ }$ System \ref{e:base}, \ref{e:cons} consists of $O(n^4)$ linear equations and inequalities with $O(n^4)$ unknowns. The existence/absence of the system's solutions can be detected using the ellipsoid algorithm \cite{khach,gls}. Because all coefficients of the system are $0$ or $1$, the ellipsoid algorithm will solve this system in strongly polynomial time. 
\newline\indent
$\mbox{	}$ Let's notice that constrains \ref{e:cons} explicitly involve the input and pattern's vertex labeling trough their adjacency matrices - see definition \ref{e:def} of the compatibility boxes. Thus, system \ref{e:base}, \ref{e:cons} is an asymmetric linear system. It can be seen that the system's solutions constitute a convex subset of the Birkhoff polytope \cite{bir} in $R^{n^4}$. Vertex relabeling of digraphs $G$ and $S$ will rotate that subset all over the polytope.

\section{Examples}
\label{s:xmp}

$\mbox{	}$ Let's use our method and resolve the following SubGI instances.
\begin{description}
\item[Vertex vs vertex:]
Let input and pattern have just one vertex each:
\[
A_G = (g_{1,1})_{1 \times 1}, ~ A_S = (s_{1,1})_{1 \times 1}
\]
System \ref{e:base} for $n=1$ looks as follows:
\[
y_{1,1,1,1} = 1
\]
Constrains \ref{e:cons} for the instance look as follows:
\[
y_{1,1,1,1} = \left \{ \begin{array}{cl}
1, & s_{1,1} \leq g_{1,1} \\
0, & s_{1,1} > g_{1,1} \\
\end{array} \right.
\]
Thus, the resolution for this SubGI instance is ``YES'' iff  there is no excess of loops in the pattern:
\[
s_{1,1} \leq g_{1,1}
\]
\item[Arc vs arc:]
Let input and pattern be just arcs:
\[
G: 1\rightarrow 2, ~ S: 1\leftarrow 2
\]
For $n=2$, system \ref{e:base} looks as follows:
\begin{equation}
\label{e:n2}
\begin{array}{rcl}
x_{1,2,1,2} & = & y_{2,2,2,2}  \\
x_{1,2,2,1} & = & y_{2,2,1,1}  \\
x_{2,1,1,2} & = & y_{1,1,2,2}  \\
x_{2,1,2,1} & = & y_{1,1,1,1}  \\
y_{1,1,1,1} + y_{1,1,2,2} & = & 1 \\
y_{2,2,1,1} + y_{2,2,2,2} & = & 1 \\
\end{array}
\end{equation}
Constrains \ref{e:cons} for the given input and pattern may be presented as follows:
\[
y_{1,1,1,1} = 0, ~ y_{2,2,2,2} = 0
\]
The aggregated system has a solution:
\[
\begin{array}{l}
x_{1,2,2,1}  =  y_{2,2,1,1} = x_{2,1,1,2}  =  y_{1,1,2,2} = 1 \\
x_{1,2,1,2}  =  y_{2,2,2,2} = x_{2,1,2,1}  =  y_{1,1,1,1} = 0 \\
\end{array} 
\]
Thus, the resolution for the given SubGI instance is ``YES''. The appropriate relabeling of the pattern is transposition $(1,2)$.
\item[Arc vs loop:]
Let input and pattern be an arc and a loop appropriately:
\[
G: 1\rightarrow 2, ~ S: 1 \rightarrow 1
\]
Adding to $S$ one isolated vertex with index 2 will produce the case of $n=2$. System \ref{e:base} for the case is system \ref{e:n2}, and constrains \ref{e:cons} for the instance may be presented as follows:
\[
y_{1,1,1,1} = 0, ~ y_{1,1,2,2} = 0
\]
The aggregated system has no solutions:
\[
1 = y_{1,1,1,1} + y_{1,1,2,2} = 0
\]
Thus, the resolution for the given SubGI instance is ``NO''.
\item[Arc/loop vs loop/arc:]
Let input and pattern be the following digraphs:
\[
G: 1\rightarrow 2 \rightarrow 2, ~ S: 1 \rightarrow 1 \rightarrow 2
\]
System \ref{e:base} for the case is system \ref{e:n2}, and constrains \ref{e:cons} for the instance may be presented as follows:
\[
y_{1,1,1,1} = 0, ~ x_{1,2,2,1} = 0
\]
The aggregated system has no solutions:
\[
1 = y_{1,1,2,2} = x_{1,2,2,1} = 0
\]
Thus, the resolution for the given SubGI instance is ``NO''.
\item[Edge vs arc:]
Let input and pattern be the following digraphs:
\[
G: 1\rightarrow 2 \rightarrow 1, ~ S: 1 \rightarrow 2
\]
System \ref{e:base} for the case is system \ref{e:n2}, and there are no constrains \ref{e:cons} for the instance. Thus, the aggregated system consists of system \ref{e:n2} alone. The center of its solutions is the following point:
\[
\forall i,j,\mu,\nu ~ (x_{ij\mu\nu} = y_{ii\mu\mu} = 1/2)
\]
Thus, the resolution for the given SubGI instance is ``YES''.
\item[Cycle vs edge:]
Let input and pattern be the following digraphs:
\[
G: 1\rightarrow 2 \rightarrow 3 \rightarrow 1, ~ S: 1 \rightarrow 2 \rightarrow 1
\]
Compatibility matrix $C$ for this SubGI instance looks as follows:
\[
\begin{array}{|ccc|ccc|ccc|}
\hline
1&0&0 & 0&0&0 & 0&1&1 \\
0&1&0 & 0&0&0 & 1&0&1 \\
0&0&1 & 0&0&0 & 1&1&0 \\
\hline
0&0&0 & 1&0&0 & 0&1&1 \\
0&0&0 & 0&1&0 & 1&0&1 \\
0&0&0 & 0&0&1 & 1&1&0 \\
\hline
0&1&1 & 0&1&1 & 1&0&0 \\
1&0&1 & 1&0&1 & 0&1&0 \\
1&1&0 & 1&1&0 & 0&0&1 \\
\hline
\end{array}
\]
Compatibility boxes entirely filled with $0$ will produce constrains \ref{e:cons} incompatible with system \ref{e:base}, i.e. the aggregated system \ref{e:base} and \ref{e:cons} will have no solutions. Thus, the resolution for the given SubGI instance is ``NO''.
\item[Cycle vs path:]
Let input and pattern be the following digraphs:
\[
G: 1\rightarrow 2 \rightarrow 3 \rightarrow 1, ~ S: 1 \rightarrow 2 \rightarrow 3
\]
Compatibility matrix $C$ for this SubGI instance looks as follows:
\[
\begin{array}{|ccc|ccc|ccc|}
\hline
\it{1}&0&0 & 0&\it{1}&0 & 0&1&\it{1} \\
0&1&0 & 0&0&1 & 1&0&1 \\
0&0&\bf{1} & \bf{1}&0&0 & 1&\bf{1}&0 \\
\hline
0&0&\bf{1} & \bf{1}&0&0 & 0&\bf{1}&0 \\
\it{1}&0&0 & 0&\it{1}&0 & 0&0&\it{1} \\
0&1&0 & 0&0&1 & 1&0&0 \\
\hline
0&1&1 & 0&0&1 & 1&0&0 \\
1&0&\bf{1} & \bf{1}&0&0 & 0&\bf{1}&0 \\
\it{1}&1&0 & 0&\it{1}&0 & 0&0&\it{1} \\
\hline
\end{array}
\]
Constrains \ref{e:cons} produced by this compatibility matrix are compatible with system \ref{e:base}, i.e. the aggregated system has solutions. Two of the three solution grids of the system are shown in the above matrix in italic and in bold. Thus, the resolution for the given SubGI instance is ``YES''.
\item[Cycle vs cycle:]
Let input and pattern be the following digraphs:
\[
G: 1\rightarrow 2 \rightarrow 3 \rightarrow 4 \rightarrow 1, ~ S: 1 \rightarrow 2 \rightarrow 3 \rightarrow 1
\]
Compatibility matrix $C$ for this SubGI instance looks as follows:
\[
\begin{array}{|cccc|cccc|cccc|cccc|}
\hline
1&0&0&0 & 0&1&0&0 & 0&0&0&1 & 0&1&1&1 \\
0&1&0&0 & 0&0&1&0 & 1&0&0&0 & 1&0&1&1 \\
0&0&1&0 & 0&0&0&1 & 0&1&0&0 & 1&1&0&1 \\
0&0&0&1 & 1&0&0&0 & 0&0&1&0 & 1&1&1&0 \\
\hline
0&0&0&1 & 1&0&0&0 & 0&1&0&0 & 0&1&1&1 \\
1&0&0&0 & 0&1&0&0 & 0&0&1&0 & 1&0&1&1 \\
0&1&0&0 & 0&0&1&0 & 0&0&0&1 & 1&1&0&1 \\
0&0&1&0 & 0&0&0&1 & 1&0&0&0 & 1&1&1&0 \\
\hline
0&1&0&0 & 0&0&0&1 & 1&0&0&0 & 0&1&1&1 \\
0&0&1&0 & 1&0&0&0 & 0&1&0&0 & 1&0&1&1 \\
0&0&0&1 & 0&1&0&0 & 0&0&1&0 & 1&1&0&1 \\
1&0&0&0 & 0&0&1&0 & 0&0&0&1 & 1&1&1&0 \\
\hline
0&1&1&1 & 0&1&1&1 & 0&1&1&1 & 1&0&0&0 \\
1&0&1&1 & 1&0&1&1 & 1&0&1&1 & 0&1&0&0 \\
1&1&0&1 & 1&1&0&1 & 1&1&0&1 & 0&0&1&0 \\
1&1&1&0 & 1&1&1&0 & 1&1&1&0 & 0&0&0&1 \\
\hline
\end{array}
\]
Constrains \ref{e:cons} produced by this compatibility matrix are incompatible with system \ref{e:base}, i.e. the aggregated system has no solutions. To see that, let's apply system \ref{e:base} to the compatibility matrix as constrains on its elements. To satisfy these constrains at least partially, the forth box column and the forth box row of the compatibility matrix have to be trimmed/depleted as follows:
\[
\begin{array}{|cccc|cccc|cccc|cccc|}
\hline
1&0&0&0 & 0&1&0&0 & 0&0&0&1 & 0&0&1&0 \\
0&1&0&0 & 0&0&1&0 & 1&0&0&0 & 0&0&0&1 \\
0&0&1&0 & 0&0&0&1 & 0&1&0&0 & 1&0&0&0 \\
0&0&0&1 & 1&0&0&0 & 0&0&1&0 & 0&1&0&0 \\
\hline
0&0&0&1 & 1&0&0&0 & 0&1&0&0 & 0&0&1&0 \\
1&0&0&0 & 0&1&0&0 & 0&0&1&0 & 0&0&0&1 \\
0&1&0&0 & 0&0&1&0 & 0&0&0&1 & 1&0&0&0 \\
0&0&1&0 & 0&0&0&1 & 1&0&0&0 & 0&1&0&0 \\
\hline
0&1&0&0 & 0&0&0&1 & 1&0&0&0 & 0&0&1&0 \\
0&0&1&0 & 1&0&0&0 & 0&1&0&0 & 0&0&0&1 \\
0&0&0&1 & 0&1&0&0 & 0&0&1&0 & 1&0&0&0 \\
1&0&0&0 & 0&0&1&0 & 0&0&0&1 & 0&1&0&0 \\
\hline
0&0&1&0 & 0&0&1&0 & 0&0&1&0 & 1&0&0&0 \\
0&0&0&1 & 0&0&0&1 & 0&0&0&1 & 0&1&0&0 \\
1&0&0&0 & 1&0&0&0 & 1&0&0&0 & 0&0&1&0 \\
0&1&0&0 & 0&1&0&0 & 0&1&0&0 & 0&0&0&1 \\
\hline
\end{array}~
\]
After this depletion, the fact that the fourth box column contradicts with the third group of equations in system \ref{e:base} becomes obvious. Thus, the resolution for the given SubGI instance is ``NO''.
\end{description}

\section{Linear program for TSP}
\label{s:tsp}

$\mbox{	}$ Let input $G$ be an arc-weighted digraph, i.e. let each arc in $G$ have a weight. TSP is a problem of finding a Hamiltonian cycle in $G$ with the minimal total weight\footnote{Because $G$ is a digraph, we actually consider here  the Asymmetric Traveling Salesman Problem (ATSP).}. That is a NP-complete problem \cite{karp}.
\begin{description}
\item[~~~~]

$\mbox{	}$ The pattern $S$ for TSP is any circular permutation matrix, for example:
\[
S = \left ( \begin{array}{ccccc}
0 & 0 & 0 & \ldots & 1 \\
1 & 0 & 0 & \ldots & 0 \\
0 & 1 & 0 & \ldots & 0 \\
\vdots&\ddots&\ddots&\ddots&\vdots \\
\end{array} \right )_{n\times n}
\]
Let's construct system \ref{e:cons} for SubGI instance $(G,S)$. Then, aggregated linear system \ref{e:base} and \ref{e:cons} will express the Hamiltonian Cycle Problem which is a NP-complete problem \cite{karp}, as well.
\newline
$\mbox{	}$ Let $w(\mu,\nu)$ be a weight function - the weight of the arc from vertex $\mu$ into vertex $\nu$ in input $G$. As usual, let $w(\mu,\nu) = + \infty$ for non-adjacent vertices. Then, the following asymmetric polynomial size linear program will express TSP:
\[
\sum_{i,j,\mu,\nu} w(\mu,\nu) x_{ij \mu\nu} ~ \rightarrow ~ \min
\] 
- subject to constrains \ref{e:base} and \ref{e:cons}.
\newline
$\mbox{	}$ From the practical perspective, let's notice that we do not require function $w(\mu,\nu)$ to be positive.
\end{description}

\section{Linear model for SAT}
\label{s:sat}

\begin{description}
\item[~~~~]

$\mbox{ }$ In 1971, Cook \cite{cook0} found that with a polynomial number of operations any non-deterministic Turing machine (NDTM) can be expressed by the appropriate conjunctive normal form (CNF): the question of whether there is an acceptable input is a question of whether the appropriate CNF is satisfiable. That made SAT the first NP-complete problem, because it is a NP-problem and the very words ``NP-problem'' mean a problem which can be solved by NDTM in polynomial time. In 1973, Levin \cite{levin} independently repeated the result in terms of search. In 1972, Karp \cite{karp} selected SAT as a root of NP-completeness theory: a problem is NP-complete if SAT can be reduced to that problem in polynomial time, and visa versa.
\newline
$\mbox{ }$ Let $f$ be a given CNF:
\[
f = c_1 \wedge c_2 \wedge \ldots \wedge c_m
\] 
- where clause $c_i$ is a disjunction of $k_i$ literals - some Boolean variables or their negations. Formula $f$ defines an instance of SAT: whether there is such a true-assignment to the involved Boolean variables which would make $f = true$.
\newline
$\mbox{ }$ Ultimately, we could apply the distributive laws and rewrite formula $f$ in a disjunctive form (DF). That would reduce SAT to an existence problem for implicants in the emerging DF. This last problem can be easily expressed as a SubGI instance.
\newline
$\mbox{ }$ Let's enumerate literals in each of the clauses in formula $f$. For each couple of clauses $(c_i,c_j)$, let's build a compatibility box: the $(\alpha,\beta)$-element in the matrix is $0$ or $1$ depending on whether the $\alpha$-th literal in clause $c_i$ and the $\beta$-th literal in clause $c_j$ are complimentary. Let's aggregate all these compatibility boxes in a box matrix. Obviously, there is an implicant in the DF of $f$ iff there is a grid of elements in the box matrix, one element per compatibility box, whose all elements are equal $1$. Each such grid of elements consists of the couples of literals which participate in an implicant.
\newline
$\mbox{ }$ The box matrix built in such a way may be seen as input $G$. Then, pattern $S$ may be a box matrix of the same structure as $G$ but whose boxes are entirely filled with $0$ except their upper-left-corner elements, which are equal $1$:
\[
S = (S_{ij})_{m\times m}, ~ S_{ij} = \left ( \begin{array} {ccc}
1&0&\ldots \\
0&0&\ldots \\
\vdots&\vdots&\ddots \\
\end{array} \right )_{k_i \times k_j}
\]
There is one obvious restriction on the relabeling of $S$: the elements of boxes $S_{ij}$ are not allowed to leave their boxes. This restriction can be accommodated in system \ref{e:cons} with a polynomial number of additional linear constrains.

\end{description}

\section*{~~~ Conclusion}

\begin{description}
\item[~~~~]

$\mbox{	}$ We described a polynomial time reduction of SubGI to a polynomial size asymmetric linear system. The system consists of systems \ref{e:base} and \ref{e:cons}. Subsystem \ref{e:base} depends on the size of SubGI instance, only. Subsystem \ref{e:cons} describes the structure of the given input and pattern. The system's asymmetry is due to the explicit involvement of the input and pattern's adjacency matrices in the construction of system \ref{e:cons} - see definition \ref{e:def}. So, the result may be seen as complimentary to the Yannakakis theorem \cite{yan}. 
\newline
$\mbox{ }$ Linear system \ref{e:base}, \ref{e:cons} defines a sub-polytope in the Birkhoff polytope. Vertices of this sub-polytope are those permutation matrices which satisfy quadratic integral system \ref{e:quadratic}. Relabeling of the input and pattern rotates this sub-polytope all over the Birkhoff polytope.
\newline
$\mbox{ }$ Ultimately, system \ref{e:base}, \ref{e:cons} may be seen as a parallel testing of all guesses, where guesses are $n\times n$ permutation matrices - the unknowns in system \ref{e:quadratic}. Basically, this parallelization was achieved with encoding SubGI in the contradictions between relabeling possibilities for different vertices.
\newline
$\mbox{ }$ Obviously, the described ``continuous'' solution of SubGI is not unique. Also, we could develop a polynomial time discrete algorithm which would search the compatibility matrix for the solution grids as, for example, it was done in \cite{3sat} for 3SAT which is a NP-complete problem \cite{cook0}, too\footnote{For 3SAT, see a demo at http://www.timescube.com}.
\end{description}

\end{document}